\documentclass[10pt]{article}

\usepackage{amssymb}
\usepackage{graphicx}
\usepackage[numbers,compress]{natbib}
\begin{document}

	\title{Conserved spin operator of Dirac's theory in spatially flat FLRW spacetimes}
	
	\author{Ion I. Cot\u aescu\thanks{Corresponding author E-mail:~~i.cotaescu@e-uvt.ro}\\
	{\it West University of Timi\c soara,} \\{\it V. Parvan Ave. 4,
		RO-300223 Timi\c soara}}

\maketitle
\begin{abstract}
New conserved spin and orbital angular momentum operators  of Dirac's theory on spatially flat FLRW spacetimes are proposed generalizing thus  recent results concerning the role of Pryce's spin operator in the flat case   [I. I. Cot\u aescu, Eur. Phys. J. C  (2022) 82:1073]. These operators split the conserved total angular momentum generating the new spin and orbital symmetries that form  the rotations of the isometry groups. The new spin operator is defined and  studied in active mode with the help of a suitable spectral representation giving its Fourier transfor. Moreover, in the same manner is defined the operator of the fermion polarization. The orbital angular momentum is derived in passive mode using a new method, inspired by Wigner's theory of induced representations, but working properly only for global rotations. In this approach the quantization is performed finding that the one-particle spin and orbital angular momentum operators have the same form in any FLRW spacetime regardless their concrete geometries given by various scale factors. 
\end{abstract}

\section{Introduction}

The spin operator is defined naturally in the non-relativistic quantum mechanics where this is conserved commuting with  the Hamiltonian operator. In contrast, the would-be spin operator of Dirac's theory in the relativistic quantum mechanics (RQM) whose components are the rotation generators of the $SL(2,\mathbb{C})$ group does not commute with the   the Dirac Hamiltonian neither in Minkowski flat spacetime nor in spatially flat Friedman-Lema\^ itre-Robertson Walker (FLRW) manifolds. For this reason many authors studied the possibility of defining new conserved spin and orbital angular momentum operators splitting properly the conserved operator of total angular momentum. 

This problem was solved in Minkowski spacetime  by Pryce which proposed the desired conserved spin operator in momentum representation (MR) giving its Fourier transform many time ago. As this result was written in the active mode in which the operators act on the mode spinors of MR it was less relevant being ignored for more than seven decades. Recently, by using a new spectral representation of the Fourier operators we  re-defined this operator in configuration representation (CR) showing that this is just the spin operator we need in Dirac's theory on Mikowski spacetime \cite{Cot}.  Moreover, in the same manner we defined the operator of fermion polarization in the most general case when the direction of spin projections depends on momentum. In other respects, for avoiding the difficulties in manipulating operators in active mode we proposed the method of associated operators in passive mode relating any operator acting on the Dirac field in CR to a pair of operators acting directly on the wave spinors in MR \cite{Cotnew}.  In this framework we studied the entire operator algebra of Dirac's theory in special relativity performing then the quantization giving the conserved one-particle operators as well as the operators having oscillating terms as those producing zitterbewegung \cite{Cotnew}.

In the present paper we would like to extend this study to any spatially flat FLRW spacetime but restricting ourselves only to the conserved operators generating the spin and orbital symmetries, i. e. the  new conserved spin and orbital angular momentum operators we have to study here.   For this purpose we use both the spectral representations in active mode and the method of associated operators in passive mode. In Minkowski spacetime these methods are based on the special structure of the mode spinors constructed according to Wigner's theory of induced representations \cite{Wig,Mc,Wa,Th,WKT}.  As this theory cannot be applied in the case of FLRW spacetimes where we do not have boost transformations, we are forced to use an alternative  approach but which works properly only for global rotations.  In this manner we derive the associated isometry generators pointing out the associated spin and orbital angular momentum operators we need for applying the Bogolyubov method of quantization \cite{Bog}.  Finally we find that the conserved one-particle operators of the quantum field theory (QFT) have the same forms in any FLRW spacetime as being independent on time and implicitly on the scale factors.  In this paper we use exclusively conformal frames with conformal coordinates and diagonal tetrad gauge.

We start in the next section presenting the Lagrangian theory of Dirac's free field on FLRW spacetimes defining the new spin and orbital symmetries generated by the conserved spin and orbital angular momentum operators. The third section is devoted to the structure of the mode spinors in MR separating the time modulation functions governing their time evolution from the spin parts constructed by using a new method inspired  by Wigner's one applied in the flat case. We specify that these mode spinors are defined in rest frames only if we set the rest frame vacuum we proposed recently \cite{Crfv}. In the next section we adopt the active mode in which the operators act on the mode spinors in CR. Consequently, for deriving  the action of the components of the new spin operator  we must resort to our spectral representation for deriving their Fourier transforms finding how these depend on the time modulation functions. However,  in spite of this apparent time dependence the spin operator is conserved via Noether's theorem. Moreover, in Minkowski spacetime the Fourier transforms of the spin components become just those proposed by Pryce in MR \cite{B}. 

Observing then the difficulties in deriving the above results we abandon the active mode developing in Sec.  5 the mentioned method of associated operators in passive mode. This gives us the opportunity of pointing out that the covariant representation of the Dirac field in CR is equivalent with a direct product of representations carried by the wave spinors in MR, just as in Wigner's theory in Minkowski spacetime.  Hereby we derive the associated operators to the isometry generators preparing the quantization procedure performed in Sec. 6 where we derive the corresponding one-particle operators of QFT which have the same expressions in all the FRLW spacetimes including the Minkowski one. Finally we present our concluding remarks. Some technical details related to the $SL(2,\mathbb{C})$ are presented in an Appendix.   

\section{Covariant Dirac field on spatially flat FLRW spacetimes}

The Dirac free field on $(1+3)$-dimensional local Minkowskian  spacetimes $M$ may be defined in frames $\{x;e\}$ formed by local chats, $\{x\}$, and orthogonal frames defined by the terades, $e$. The  coordinates  $x^{\mu}$ are labeled by natural inidices, $\alpha,...\mu,\nu,...=0,1,2,3$ while the vector fields $e_{\hat\alpha}=e_{\hat\alpha}^{\mu}\partial_{\mu}$ defining the local orthogonal frames,  and the 1-forms $\omega^{\hat\alpha}=\hat e_{\mu}^{\hat\alpha}dx^{\mu}$ of the dual coframes  are labeled by local indices, $\hat\mu,\hat\nu,...=0,1,2,3$. The natural indices are raised or lowered  by the metric  tensor of $M$,  $g_{\mu\nu}=\eta_{\hat\alpha\hat\beta}\hat e^{\hat\alpha}_{\mu}\hat e^{\hat\beta}_{\nu}$, while for the local ones we have to use  the Minkowski  metric  $\eta={\rm diag}(1,-1,-1,-1)$. In an arbitrary frame $\{x;e\}$ of  $M$ the tetrad-gauge invariant action of the Dirac field $\psi : M\to {\cal V}_D$, of mass $m$, minimally coupled to background's gravity, reads,  
\begin{equation}\label{action}
	{\cal S}[e,\psi]=\int\, d^{4}x\sqrt{g}\left\{
	\frac{i}{2}[\overline{\psi}\gamma^{\hat\alpha}\nabla_{\hat\alpha}\psi-
	(\overline{\nabla_{\hat\alpha}\psi})\gamma^{\hat\alpha}\psi] -
	m\overline{\psi}\psi\right\}
\end{equation}
where  $\bar{\psi}=\psi^+\gamma^0$ is the Dirac adjoint of $\psi$ and  $g=|\det(g_{\mu\nu})|$. 
The field $\psi$ takes values in the space of Dirac spinors ${\cal V}_D$ which carries the Dirac representation $\rho_D= (\frac{1}{2},0)\oplus (0,\frac{1}{2})$ of the $SL(2,{\Bbb C})$ group. In this representation one can define the Dirac matrices $\gamma^{\hat\alpha}$ (with local indices) which are Dirac self-adjoint, $\overline{\gamma^{\hat\mu}}=\gamma^0{\gamma^{\hat\mu}}^+\gamma^0=\gamma^{\hat\mu}$, and satisfy the anti-commutation rules (\ref{acom}). These matrices may extend the $sl(2,{\Bbb C})$ Lie algebra to a  $su(2,2)$ one such that $\rho_D$ is in fact much larger than a representation of the $SL(2,{\Bbb C})$ group. The gauge invariant field theories use covariant derivatives of the form
\begin{equation}\label{cov}
	\nabla_{\hat\alpha}=e^{\mu}_{\hat\alpha}\nabla_{\mu}= e_{\hat\alpha}+\frac{i}{2}\hat\Gamma^{\hat\gamma}_{\hat\alpha
		\hat\beta}s^{\hat\beta\,
		\cdot}_{\cdot\, \hat\gamma}\,,
\end{equation}
where $s^{\hat\alpha \hat\beta}$ are the $SL(2,{\Bbb C})$ generators (\ref{gen})  while 
$\hat\Gamma^{\hat\sigma}_{\hat\mu \hat\nu}=e_{\hat\mu}^{\alpha} e_{\hat\nu}^{\beta}
(\hat e_{\gamma}^{\hat\sigma}\Gamma^{\gamma}_{\alpha \beta} -\hat e^{\hat\sigma}_{\beta, \alpha})$ are the connection components in local frames  expressed in terms of tetrads and Christoffel symbols,   $\Gamma^{\gamma}_{\alpha \beta}$. Note that these connections are known as the spin connections and denoted often by $\Omega$. The covariant derivatives of the Dirac operator $E_x=  i\gamma^{\hat\alpha}\nabla_{\hat\alpha}-m$ guarantee that the Dirac equation $E_{x}\psi=0$  is {\em gauge invariant} in the sense that this does not change its form when we change the positions of local frames performing gauge transformations.   The particular solutions of the Dirac equation form a vector space that may be equipped with the  relativistic scalar product  \cite{Crfv}
 \begin{equation}\label{SPD}
 	\langle \psi, \psi'\rangle_D=\int_{\Sigma} d\sigma_{\mu} \sqrt{g}\, e^{\mu}_{\hat\alpha}
 	\,\bar{\psi}(x)\gamma^{\hat\alpha}\psi'(x)\,, 
 \end{equation}
 whose integral is performed on a space-like section $\Sigma\subset M$. 
 
  In what follows we study the Dirac free field on $(1+3)$-dimensional spatially flat FLRW spacetimes $M(a)$ having scale factors $a$.   We work exclusively in conformal frames $\{x_c;e\}$ formed by conformal chats, $\{t_c,{\vec x}_c\}$, and orthogonal frames defined by terades, $e$.  The conformal coordinates  are the conformal time $t_c$ and the co-moving Cartesian space coordinates, $x_c^i$  ($i,j,k...=1,2,3$), giving the line element
 \begin{eqnarray}
 	ds^2=g_{\mu\nu}(x_c)dx_c^{\mu}dx_c^{\nu}=a(t_c)^2(dt_c^2-d{\vec x}_c\cdot d{\vec x}_c)\,.
 \end{eqnarray}
 Here we adopt the diagonal tetrad-gauge in which the local frames are defined by   
 \begin{eqnarray}
 	e_0&=&\frac{1}{a(t_c)}\,\partial_{t_c}\,,\qquad  \omega^0=a(t_c)dt_c\,, \\
 	e_i&=&\frac{1}{a(t_c)}\,\partial_i\,, \qquad ~~ \omega^i=a(t_c)dx^i\,,
 \end{eqnarray}
 such that we preserve the {global} $SO(3)$ symmetry  allowing us to use systematically the  $SO(3)$ vectors.  In these frames the Dirac equation can be put in Hamiltonian form, $i\partial_{t_x}\psi=H_c\psi$, with the help of the Hamiltonian operator 
\begin{equation}\label{Hc}
H_c({x_c})=m a(t_c)\gamma^0- i\gamma^0\gamma^i\partial_{x^i_c}
	-\frac{3i}{2}\frac{\dot{a}(t_c)}{a(t_c)}\,,\quad \dot a(t_c)=\frac{d a(t_c)}{dt_c}\,.
\end{equation} 
while the scalar product takes the form
\begin{equation}\label{spD}
	\langle \psi, \psi'\rangle_D=\int_{\Sigma} d^{3}x_c\,
	a(t_c)^{3}\bar{\psi}(x_c)\gamma^{0}\psi(x_c)=\int_{\Sigma} d^{3}x_c\,
	a(t_c)^{3}{\psi}^+(x_c)\psi(x_c)\,, 
\end{equation} 
where the integration is performed over a flat space section, $\Sigma=\mathbb{R}^3$. 

In the spacetimes $M(a)$ the action (\ref{action})  is invariant at least under {\em  global} isometries, $(R,a): x_c^i\to x_c^{i\,\prime}=R^{i\,\cdot}_{\cdot\, j}x_c^j+a^i$ formed by rotations $R\in SO(3)$ and three-dimensional translations, ${\vec a}\in T(3)$. The isometry group $E(3)=T(3)\circledS SO(3)$   is a semidirect product  where $T(3)$ is the invariant subgroup.   The Dirac field transforms   under isometries  according to the {\em covariant} representation  ${T}   \,:\,(r, {\vec a})\to { T}_{r,{\vec a}}$ of the  group $\tilde E(3)=T(3)\circledS SU(2)$ which is the universal covering group of the group $E(3)$ \cite{WKT}. We have thus the global transformations 
\begin{equation}\label{TAa}
	({T}_{r,{\vec a}}\psi  )(t_c,{\vec x}_c)
	=r \psi  \left(t_c, R(\hat r)^{-1}({\vec x}_c-{\vec a})\right)\,, \quad \forall r\in \rho_D[SU(2)]\,,\, {\vec a}\in T(3).
\end{equation}
generated by  the basis-generators of the  corresponding representation of  the algebra ${\rm Lie}({T})$ that read
\begin{eqnarray}\label{Pgen}
	P^{i}=\left.i\frac{\partial {T}_{1,a}}{\partial a^{i}}\right|_{a=0}\,, \quad 
	J_{i}=\left.i\frac{\partial {T}_{r(\theta),0}}{\partial \theta^i}\right|_{\theta=0}\,,
\end{eqnarray}
where $r(\theta)$ is defined in Eq. (\ref{r}). We obtain thus the momentum components,  $ P^i=-i\partial_{x_c^i}$, and those of the total angular momentum,  
\begin{eqnarray}
	J_i  &=&\frac{1}{2}\,\varepsilon_{ijk}	J  _{jk}=
	-i\varepsilon_{ijk}\underline{x}^j\partial_{x_c^k}+s_i  \,,\label{J}
\end{eqnarray}
where the components $\underline{x}^i$ of  the  position (or coordinate) vector-operator $\underline{\vec x}$  act as $(\underline{x}^i \psi)(x_c)=x_c^i\psi(x_c)$ while the reducible matrices $s_i$  are given by Eq. (\ref{si}). The operators $P^i$ and $J_i$ are the standard basis-generators of the algebra ${\rm Lie}(T)$. As the action (\ref{action}) is  invariant under isometries the scalar product (\ref{spD}) is also  invariant,  
\begin{equation}
	\langle {T}_{\lambda,a}\psi  ,{T}_{\lambda,a}\psi'  \rangle_D=\langle\psi ,\psi' \rangle_D\,.	
\end{equation}
The generators $X\in {\rm Lie}(T)$  are self-adjoint operators, $ \langle\psi, X\psi'  \rangle_D =\langle X\psi  ,\psi'  \rangle_D$,  conserved via Noetrher's theorem. 
Therefore, we may conclude that in this framework the covariant representation ${T}$ behaves as a {\em unitary} one with respect to the relativistic scalar product  (\ref{spD}).

The subgroup $SU(2)\subset\tilde E(3)$  is of a special interest  such that we denote by $T_{r,0}=T^r_{\hat r}$ the restriction of the covariant representation ${T}$ to this subgroup.  The basis-generators of the representation $T^r$  are the components of  total angular momentum operator, ${\vec J}=\underline{\vec x}\land{\vec P} +{\vec s}$, defined by Eq. (\ref{J}), which is formed by the orbital term $\underline{\vec x}\land{\vec P}$ and  the  spin matrix ${\vec s}\in sl(2,\mathbb{C})$. As these operators are not conserved separately via Noetrher's theorem we must look for a new conserved spin operator ${\vec S}$ associated to a suitable new position operator, ${\vec X}=\underline{\vec x}+\delta{\vec X}$, allowing the  splitting
\begin{equation}\label{spli}
	{\vec J}=\underline{\vec x}\land{\vec P} +{\vec s}={\vec L}+{\vec S}	\,,\qquad {\vec L}={\vec X}\land {\vec P}\,,
\end{equation}
which impose  the correction $\delta{\vec X}$ to satisfy $\delta{\vec X}\land {\vec P}={\vec s}-{\vec S}$. This new splitting gives rise to a pair of  new $su(2)\sim so(3)$ symmetries, namely the {\em orbital} symmetry generated by $\{L_1,L_2,L_3\}$ and the {\em spin} one generated by $\{S_1,S_2,S_3\}$.  

The plane wave solutions of the Dirac equation depend on {\em arbitrary} Pauli spinors  $\xi=\{\xi_{\sigma}| \sigma=\pm\frac{1}{2}\}$ determining the fermion polarization. These spinors form similar bases in both the spaces  ${\cal V}_P$ of Pauli spinors carrying the irreducible representations $(\frac{1}{2},0)$ and $(0,\frac{1}{2})$ of $\rho_D$. The basis  of polarization spinors can be changed at any time, $\xi\to \hat r\xi$, by  applying a $SU(2)$ rotation $\hat r$. For this reason, when we study this symmetry  it is convenient to denote the Dirac field by $\psi_{\xi}$ instead of $\psi$  pointing out explicitly its dependence on the polarization spinors. We define now the transformations of spin symmetry with the help of a representation ${T}^s:\hat r\to {T}^s_{\hat r}$ of the group $SU(2)$ acting as,   
\begin{equation}\label{Rs}
	\left(	T^s_{
	\hat r(\theta)}\psi_{\xi}\right)(x_c)=\psi_{\hat r(\theta)\xi}(x_c)\,,
\end{equation} 
where $\hat r(\theta)$ are the rotations (\ref{r}) with Cayley-Klein parameters. Then the components of the spin operator can be defined as the generators of this representation, \cite{Cot}
\begin{equation}\label{Spipi}
	S_i=\left.i\frac{\partial T^s_{\hat r(\theta)}}{\partial \theta^i}\right|_{\theta^i=0} 	~~\Rightarrow~~ S_i\psi_{\xi}=\psi_{\hat s_i\xi}\,,
\end{equation}
whose action is obvious. Similarly we define   the orbital representation $T^o: \hat r\to T^o_{\hat r}$ as
\begin{equation}\label{Ro}
	\left(T^o_{\hat r(\theta)}\psi_{\xi}\right)(t_c,{\vec x}_c)=r(\theta)\psi_{\hat r(\theta)^{-1}\xi}\left(t_c, R[\hat r(\theta)]^{-1}{\vec x}_c \right)\,,
\end{equation}
for accomplishing the factorization ${T}^r=T^o \otimes T^s$. The  basis-generators of the orbital representation,
\begin{equation}\label{Lpipi}
	L_i=\left.i\frac{\partial T^o_{\hat r(\theta)}}{\partial \theta^i}\right|_{\theta^i=0} \,,
\end{equation} 
are the components of the new conserved orbital angular momentum operator. In what follows we shall pay a special attention to the new operators ${\vec S}$ and  ${\vec L}$.

\section{Mode spinors in momentum representation}

The Dirac equation  can be solved formally on $M(a)$ in conformal  frames $\{x_c;e\}$  allowing solutions of the general form    \cite{Crfv}, 
\begin{eqnarray}
	\psi({x}_c)&=&\psi^{(+)}(x_c)+\psi^{(-)}(x_c)\nonumber\\
	&=&\int d^{3}p
	\sum_{\sigma}[U_{{\vec p},\sigma}(x_c){\alpha}({\vec p},\sigma)
	+V_{{\vec p},\sigma}(x_c){\beta}^{*}({\vec p},\sigma)]\,,\label{Psi}
\end{eqnarray}
expressed in terms of  particle, $\alpha$, and antiparticle, $\beta$, wave spinors and mode spinors  $U_{{\vec p},\sigma}$  and  $V_{{\vec p},\sigma}$, of positive and respectively negative frequencies. 
These spinors satisfy the eigenvalues problems $P^i U_{\vec{p},\sigma}(t_c,{\vec x}_c)=p^i U_{\vec p,\sigma}(t_c,{\vec x}_c)$ and  $P^i V_{\vec p,\sigma}(t_c,{\vec x}_c)=-p^i V_{\vec p,\sigma}(t_c,{\vec x}_c)$ and form an orthonormal  basis being related  through the charge conjugation, 
\begin{equation}\label{chc}
	V_{{\vec p},\sigma}(t_c,{\vec x}_c)=U^c_{{\vec p},\sigma}(t_c,{\vec x}_c) =C{U}^*_{{\vec p},\sigma}(t_c,{\vec x}_c) \,, \quad C=i\gamma^2\,,
\end{equation}
satisfying the orthogonality relations
\begin{eqnarray}
	\langle U_{{\vec p},\sigma}, U_{{{\vec p}\,}',\sigma'}\rangle &=&
	\langle V_{{\vec p},\sigma}, V_{{{\vec p}\,}',\sigma'}\rangle=
	\delta_{\sigma\sigma^{\prime}}\delta^{3}({\vec p}-{\vec p}\,^{\prime})\label{ortU}\\
	\langle U_{{\vec p},\sigma}, V_{{{\vec p}\,}',\sigma'}\rangle &=&
	\langle V_{{\vec p},\sigma}, U_{{{\vec p}\,}',\sigma'}\rangle =0\,, \label{ortV}
\end{eqnarray}
with respect to the relativistic scalar product (\ref{spD}). Moreover,  this basis is supposed to be complete accomplishing the  condition  \cite{Crfv}
\begin{eqnarray}
	&&\int d^{3}p
	\sum_{\sigma}\left[U_{{\vec p},\,\sigma}(t_c,{\vec x}_c)U^{+}_{{\vec p},\sigma}(t_c,{\vec x}_c^{\,\prime})+V_{{\vec p},\sigma}(t_c,{\vec x}_c)V^{+}_{{\vec p},\sigma}(t_c,{\vec x}_c^{\,\prime})\right]\nonumber\\ &&\hspace*{32mm}=a(t_c)^{-3}\delta^{3}({\vec x}_c-{\vec x}_c^{\,\prime})\,.~~~~~\label{complet}
\end{eqnarray}
The space  of the solutions of Dirac's equation ${\cal F}=\{\psi |E_{x_c}\psi=0\}={\cal F}^+\oplus{\cal F}^-$ is formed by the subspaces of solutions of positive,  ${\cal F}^+$, and  negative, ${\cal F}^-$, frequencies which are orthogonal with respect to the scalar product (\ref{spD}). 

In RQM the physical meaning of the free field $\psi$ is encapsulated in its wave spinors
\begin{equation}\label{alpha}
	\alpha=\left( 
	\begin{array}{l}
		\alpha_{\frac{1}{2}}\\
		\alpha_{-\frac{1}{2}} 
	\end{array}\right)	\in \tilde{\cal F}^+\,,
	\quad
	\beta=\left( 
	\begin{array}{llc}
		\beta_{\frac{1}{2}}\\
		\beta_{-\frac{1}{2}} 
	\end{array}\right) 	\in \tilde{\cal F}^-\,,
\end{equation}
whose values can be obtained applying the inversion formulas
\begin{equation}\label{inv}
	\alpha_{\sigma}({\vec p})=\langle U_{{\vec p},\sigma},\psi \rangle_D\,, \quad 
	\beta_{\sigma}({\vec p})=\langle \psi,  V_{{\vec p},\sigma} \rangle_D\,, 
\end{equation}
resulted from  Eqs. (\ref{ortU}) and (\ref{ortV}). We assume now that the spaces $\tilde{\cal F}^+\sim \tilde{\cal F}^-\sim {\cal L}^2(\Omega_{\mathring{p}}, d^3p,{\cal V}_P)$ are equipped with the same scalar product, 
\begin{eqnarray}\label{spa}
	\langle \alpha, \alpha'\rangle=\int d^3p \,\alpha^+({\vec p})\alpha'({\vec p})=\int d^3p \sum_{\sigma}\alpha_{\sigma}^*({\vec p})	\alpha_{\sigma}'({\vec p})\,,
\end{eqnarray}
and similarly for the spinors $\beta$. Then after using  Eqs.  (\ref{ortU}) and (\ref{ortV}) we obtain  the important identity 
\begin{equation}\label{spp}
	\langle \psi  , \psi  '\rangle_D =\langle \alpha,\alpha'\rangle + \langle \beta,\beta'\rangle	\,,
\end{equation}
showing that  the spaces ${\cal F}$ and $\tilde{\cal F}=\tilde{\cal F}^+\oplus\tilde{\cal F}^-$ related through the expansion (\ref{Psi})  are {\em isometric}.

The general form of the mode spinors can be studied  in any manifold $M(a)$,  exploiting the Dirac equation in MR.   As here we cannot apply the Wigner method of constructing  mode spinors in Minkowski spacetime, we must look for an alternative approach inspired by the identity (\ref{id2}). For our further  purposes, it is convenient to separate from the beginning the orbital part from the spin terms assuming that the mode spinors have the general form   
\begin{eqnarray}
	U_{\vec p,\sigma}(t_c,{\vec x}_c)&=&[2\pi a(t_c)]^{-\frac{3}{2}}{e^{i{\vec p}\cdot{\vec x}_c}}{\cal U}_p(t_c)d({\vec p})u_{\sigma}({\vec p})\label{U}\\
	V_{\vec p,\sigma}(t_c,{\vec x}_c)&=&[2\pi a(t_c)]^{-\frac{3}{2}}{e^{-i{\vec p}\cdot{\vec x}_c}}{\cal V}_p(t_c)d({\vec p}) v_{\sigma}({\vec p})\label{V}
\end{eqnarray}
where we introduce the diagonal matrix-functions ${\cal U}_p(t_c)$ and ${\cal V}_p(t_c)$  which depend only on $t$ and $p=|{\vec p}\,|$, determining the time modulation of the fundamental spinors. 

The spin part is separated with the help of the Hermitian singular matrix 
\begin{equation}\label{gamp}
	d({\vec p})=d^+({\vec p})=1+\frac{\gamma^0 {\vec\gamma}\cdot {\vec p}}{p}\,, 
\end{equation}
acting on the Dirac spinors that in the standard representation of the Dirac matrices (with diagonal $\gamma^0$) have the form \cite{TH,Cot}
\begin{equation}\label{Rfspin}
	u_{\sigma}({\vec p})=\left(
	\begin{array}{c}
		\xi_{\sigma}({\vec p})\\
		0
	\end{array}\right)\,,\quad
	v_{\sigma}({\vec p})=C u^*_{\sigma}({\vec p})=\left(
	\begin{array}{c}
		0\\
		-\eta_{\sigma}({\vec p})
	\end{array}\right)\,,
\end{equation}
depending on the  Pauli spinors, $\xi_{\sigma}({\vec p})$ and $\eta_{\sigma}({\vec p})=i\sigma_2\xi^*_{\sigma}({\vec p})$ which  are normalized,  $\xi^+_{\sigma}({\vec p})\xi_{\sigma'}({\vec p})=\eta^+_{\sigma}({\vec p})\eta_{\sigma'}({\vec p})=\delta_{\sigma\sigma'}$. satisfying  the completeness condition, 
\begin{equation}\label{Pcom}
	\sum_{\sigma}\xi_{\sigma}({\vec p})\xi_{\sigma}^+({\vec p})=\sum_{\sigma}\eta_{\sigma}({\vec p})\eta_{\sigma}^+({\vec p})={\bf 1}_{2\times 2}\,.
\end{equation}
The form of these spinors depends on the direction of the spin projection which, in general, may depend on ${\vec p}$ as in the case of the helicity basis. Here we say that the polarization depending on momentum is a {\em peculiar} polarization. Otherwise, the plarization independent on momentum  will be referred as {\em common} polarization. The corresponding Dirac spinors have the completeness properties  \cite{Cot}
\begin{eqnarray}
	&&\sum_{\sigma}u_{\sigma}({\vec p}){u}^+_{\sigma}({\vec p})=\frac{1+\gamma^0}{2}\equiv\pi_+\,,\label{pip}\\
	&&\sum_{\sigma}v_{\sigma}({\vec p}){v}^+_{\sigma}({\vec p})=\frac{1-\gamma^0}{2}\equiv\pi_-\,,\label{pim}
\end{eqnarray}
laying out the Hermitian matrices $\pi_{\pm}=(\pi_{\pm})^+$ that form a complete system of projection operators as  $\pi_+\pi_-=0$ and $\pi_++\pi_-=1\in \rho_D$. All the above auxiliary quantities will be useful in the further calculations having simple calculation rules as, for example, $d({\vec p})^2=2 d({\vec p})$, $~d({\vec p})d(-{\vec p})=0$, etc..    

The principal pieces are the diagonal matrix-functions determining the time modulation of the mode spinors which can be represented as
\begin{eqnarray}
	{\cal U}_p(t_c)&=&\pi_+ u^+(t_c,p)+\pi_-u^-(t_c,p)\,,\\
	{\cal V}_p(t_c)&=&\pi_+ v^+(t_c,p)+\pi_-v^-(t_c,p)\,,
\end{eqnarray}
in terms of  the time modulation functions $u^{\pm}(t_c,p)$ and $v^{\pm}(t_c,p)$ whose  differential equations  in the general case of $m \not= 0$  may be obtained by substituting Eqs. (\ref{U}) and (\ref{V}) in Dirac's equation.  Then, after a few manipulation, we find the systems of the first order differential equations \cite{Crfv}
\begin{eqnarray}
	\left[i\partial_{t_c}\mp m\, a(t_c)\right]u^{\pm}(t_c,p)&=&{p}\,u^{\mp}(t_c,p)\,,\nonumber\\
	\left[i\partial_{t_c} \mp m\, a(t_c)\right]v^{\pm}(t_c,p)&=&-{p}\,v^{\mp}(t_c,p)\,,\label{sy2c}
\end{eqnarray}
which govern the time modulation of the free Dirac field on any spatially flat FLRW manifold. 
The solutions of this system depend on  integration constants that must be selected according to the charge conjugation condition (\ref{chc}) which requires to have   
${\cal V}_p={\cal U}^c_p=C{\cal U}_p^*C^{-1}=\gamma^5 {\cal U}_p^*\gamma^5$ 
leading to the mandatory condition
\begin{equation}\label{VU}
	v^{\pm}(t_c,p)=\left[u^{\mp}(t_c,p)\right]^*\,.
\end{equation}
The remaining normalization constants can be restricted as we have the prime integrals of the system (\ref{sy2c}), 
$\partial_{t_c} (|u^+(t_c,p)|^2+|u^-(t_c,p)|^2)=\partial_{t_c} (|v^+(t_c,p)|^2+|v^-(t_c,p)|^2)=0$, 
allowing us to impose the normalization conditions
\begin{equation}\label{uuvv}
|u^+(t_c,p)|^2+|u^-(t_c,p)|^2=|v^+(t_c,p)|^2+|v^-(t_c,p)|^2=1\,,
\end{equation}
which guarantee  Eqs.  (\ref{ortU}) and (\ref{ortV}) to be accomplished. Hereby we find  the calculation rules 
${\rm Tr}({\cal U}_p{\cal U}_p^*)={\rm Tr}({\cal V}_p{\cal V}_p^*)=2$ 
resulted from Eqs.  (\ref{uuvv}) as ${\rm Tr}(\pi_{\pm})=2$.

Gathering all the above elements we obtain the intuitive forms of the mode spinors in conformal frames and standard representation of the $\gamma$-matrices \cite{Crfv}, 
\begin{eqnarray}
U_{{\vec p},\sigma}(t_c,{\vec x}_c)	&=&\frac{e^{i{\vec p}\cdot{\vec x}_c}}{[2\pi a(t_c)]^{\frac{3}{2}}}\left(
\begin{array}{c}
u^+(t_c,p)\,	\xi_{\sigma}({\vec p})\\
u^-(t_c,p)\,\frac{\vec \sigma\cdot{\vec p}}{p}\,	\xi_{\sigma}({\vec p})
\end{array}
\right)\,,\\
V_{{\vec p},\sigma}(t_c,{\vec x}_c)	&=&-\frac{e^{i{\vec p}\cdot{\vec x}_c}}{[2\pi a(t_c)]^{\frac{3}{2}}}\left(
\begin{array}{c}
	v^+(t_c,p)\,\frac{\vec \sigma\cdot{\vec p}}{p}\,	\eta_{\sigma}({\vec p})\\
		v^-(t_c,p)\,	\eta_{\sigma}({\vec p})
\end{array}
\right)\,.
\end{eqnarray}
A special problem comes from the fact that these spinors are not defined in the rest frame where ${\vec p}=0$ because of the matrix $\frac{\vec \sigma\cdot{\vec p}}{p}$ which is undefined in this limit.  For this reason we proposed the {\em rest frame vacuum} defined by the following supplementary conditions \cite{Crfv}
\begin{equation}\label{rfv}
u^-(t_c,0)=v^+(t_c,0)=0\,, \quad 	|u^+(t_c,0)|=|v^-(t_c,0)|=1\,.
\end{equation}	
In this vacuum  the mode spinors in the rest frame are well-defined, 
\begin{eqnarray}
	U_{0,\sigma}(t_c,{\vec x})&=&[2\pi a(t_c)]^{-\frac{3}{2}}e^{-i m\, t(t_c)}u_{\sigma}(0)\,,\label{Ur}\\
	V_{0,\sigma}(t_c,{\vec x})&=&[2\pi a(t_c)]^{-\frac{3}{2}}e^{i m\, t(t_c)} v_{\sigma}(0)\,,\label{Vr}
\end{eqnarray} 
depending on the cosmic time 
\begin{equation}
	t(t_c)=\int_{t_{c\,0}}^{t_c} dt'_c a(t'_c) \,,
\end{equation}
 the rest energy of special relativity, $E_0=m$,  and the rest frame spinors $u_{\sigma}(0)$ and $v_{\sigma}(0)$. Note that in Minkowski spacetime the time modulation functions (\ref{upmM}) satisfy naturally the conditions (\ref{rfv}) defining the rest frame vacuum. In contrast, on the de Sitter expanding universe one prefers the {\em adiabatic} vacuum \cite{BD} but which cannot be defined in FLRW spacetimes with  finite big bang times. In our opinion, the de Sitter adiabatic vacuum can be seen as the partner of the rest frame one in the process of cosmological particle creation under inflation \cite{Cproc}.  

\section{Operators in active mode}

In linear algebra,  a  linear operator may act changing the basis in the active mode or transforming only the  coefficients without affecting the basis if we adopt the passive mode. 
Let us start with the operators in active mode.

The simplest operators of Dirac's theory on $M(a)$ are the multiplicative and differential operators in CR. The differential operators  are $4\times 4$ matrices depending on space derivatives, $f(i\partial_{i})\in \rho_D$, whose action on the mode spinors  
\begin{eqnarray}
	\left[f(i\partial_{i}) \psi\right](x)&=&\int d^3p \sum_{\sigma}\left[ f({p^{i}})U_{{\vec p},{\sigma}}( x)\alpha_{\sigma}({\vec p})+ f(-{p^{i}})V_{{\vec p},{\sigma}}(x)\beta^*_{\sigma}({\vec p})\right]\,,\label{F2}
\end{eqnarray}
is given by the momentum-dependent  matrices $f(p^{i})$ called the Fourier transforms of  the operators $f(i\partial_{i})$. The principal differential operators are the translation generators $P_{i}=i\partial_{i}$  and the time-dependent Hamiltonian operator (\ref{Hc}) whose Fourier transform  
\begin{equation}\label{HDp}
	H_c(t_c,{\vec p})=ma(t_c)\gamma^0+\gamma^0\vec{\gamma}\cdot {\vec p} -\frac{3i}{2}\frac{\dot{a}(t_c)}{a(t_c)} \,,
\end{equation}
allows us to write the formal actions
\begin{equation}
\left({H}_c U_{{\vec p},\sigma}\right)(x_c)=H_c(t_c,{\vec p}) U_{{\vec p},\sigma}(x_c)\,,\quad \left( {H}_cV_{{\vec p},\sigma}\right)(x_c)=H_c(t_c,-{\vec p})\,, V_{{\vec p},\sigma}(x_c)\,,	
\end{equation}	 
which will be useful for identifying conserved quantities.

 Most interesting are the integral operators whose kernels may depend on time. Here we restrict ourselves to the equal-time operators acting in conformal frames as 
\begin{equation}\label{Y1}
	(A \psi)(t_c,{\vec x}_c)=\int d^3x_c' {\frak A}(t_c,{\vec x}_c,{\vec x}_c^{\,\prime})\psi(t_c,{\vec x}_c^{\,\prime})\,,
\end{equation}
preserving  the conformal time $t_c$. The multiplication takes over this property,  
\begin{equation}
	A=A_1A_2~~ \Rightarrow ~~{\frak A}(t_c,{\vec x}_c,{\vec x}_c^{\,\prime})=\int d^3x_c''\, {\frak A}_1(t,{\vec x}_c,{{\vec x}_c}''){\frak A}_2(t_c,{{\vec x}_c}'',{\vec x}_c^{\,\prime})\,,
\end{equation}
such that the set of equal-time operators forms an algebra $E[t_c] \subset {\rm Aut}({\cal F})$ at any fixed time $t_c$. A special subalgebra $F[t_c]\subset E[t_c]$ is formed by  operators with local kernels, ${\frak A}(t_c,{\vec x}_c,{\vec x}_c')= {\frak A}(t_c,{\vec x}_c-{\vec x}_c')$, allowing  three-dimensional Fourier representations, 
\begin{equation}\label{KerY}
	{\frak A}(t_c,{\vec x}_c) =\int d^3p\,\frac{e^{i {\vec p}\cdot{\vec x}_c}}{(2\pi)^3}  {A}(t_c,{\vec p})\,,
\end{equation} 
depending on the matrices  $A(t_c,{\vec p})\in{r}_D$  we call here the Fourier transforms of the operators $A$.  Then the action (\ref{Y1}) on a field  (\ref{Psi})  can be written as
\begin{eqnarray}
&&	(A \psi)(t_c, {\vec x}_c)=	\int d^3x_c'\, {\frak A}(t,{\vec x}_c-{\vec x}_c^{\,\prime})\psi(t_c,{\vec x}_c^{\,\prime})\nonumber\\
&&~~=\int d^3p \sum_{\sigma}\left[ A(t_c,{\vec  p})U_{{\vec p},{\sigma}}(t_c, {\vec x})\alpha_{\sigma}({\vec p}) +A(t_c,-{\vec  p})V_{{\vec p},{\sigma}}(t, {\vec x})\beta^*_{\sigma}({\vec p})\right]\,.\label{Y2}
\end{eqnarray}
Remarkably,  the operator multiplication, $A=A_1A_2$, in  $F[t]$ algebra leads to the convolution of the corresponding kernels ${\frak A}={\frak A}_1*{\frak A}_2$ defined as
\begin{equation}
	{\frak A}(t_c, {\vec x}_c-{\vec x}_c^{\,\prime})=\int d^3p''{\frak A}_1(t, {\vec x}-{{\vec x}\,}''){\frak A}_2(t, {{\vec x}\,}''-{\vec x}^{\,\prime})\,,	
\end{equation} 
and, consequently, to the multiplication of their  Fourier transforms, $A(t_c,{\vec p})=A_1(t_c,{\vec p})A_2(t_c,{\vec p})$. One obtains thus the new algebra $\bar F[t_c]$ in MR formed by the Fourier transforms of the Fourier operators in which the identity is $I({\vec p})=1\in\rho _D$. Obviously, the operator $A\in F[t_c]$ is invertible if its Fourier transform is invertible in $\bar F[t_c]$. 

We say that a Fourier operator $A$ is reducible if its Fourier transform satisfies 
\begin{equation}\label{red}
	\left.\langle V_{{\vec p},{\sigma}}, A(t_c,{\vec p})U_{{\vec p},{\sigma}}\rangle_D\right|_{t_c}=	\left.\langle U_{{\vec p},{\sigma}}, A(t_c,-{\vec p})V_{{\vec p},{\sigma}}\rangle_D\right|_{t_c}=0\,,
\end{equation}
at any instant $t_c$. These operators have simple expectation values 
\begin{equation}
\left.\langle \psi, A\psi \rangle_D\right|_{t_c}= \left\langle U_{{\vec p},{\sigma}}, A(t_c,{\vec p})U_{{\vec p},{\sigma}}\rangle_D\right|_{t_c}+\left\langle V_{{\vec p},{\sigma}}, A(t_c,-{\vec p})V_{{\vec p},{\sigma}}\rangle_D\right|_{t_c}	
\end{equation}
that become conserved quantities when the Fourier transform $A(t_c,{\vec p})$ accomplishes the condition
\begin{equation}\label{cons}
\frac{d}{dt_c}\left.\langle \psi, A\psi \rangle_D\right|_{t_c}=0 ~~\Rightarrow~~ 
i\partial_{t_c}A(t_c,{\vec p})=\left[ H_c(t_c,{\vec p}), A(t_c,{\vec p})\right]\,,	
\end{equation}
allowing us to identify the conserved operators directly without resorting to Noether's theorem.

In this framework we may define the transformations of the spin symmetry (\ref{Rs}) as Fourier operators constructed with the help of the spectral representations we proposed recently. We assume that the operators $T^s_{\hat r}$ are Fourier operators whose kernels allow the spectral representation 
\begin{eqnarray}\label{kerT}
	{\frak T}^s_{\hat r}(t_c,{\vec x}_c-{\vec x}_c^{\,\prime})&=&\int d^3p\, a(t_c)^3\sum_{\sigma,\sigma'}\left[U_{{\vec p},\xi_{\sigma}}(t_c,{\vec x}_c)D_{\sigma\sigma'}(\hat r, {\vec p})U^+_{{\vec p},\xi_{\sigma'}}(t_c,{\vec x}_c^{\,\prime})\right.\nonumber\\
	&&\hspace*{20mm}\left. +V_{{\vec p},\eta_{\sigma}}(t_c,{\vec x}_c) D^{*}_{\sigma\sigma'}(\hat r, {\vec p})V^+_{{\vec p},\eta_{\sigma'}}(t_c,{\vec x}_c^{\,\prime})\right]\,,
\end{eqnarray} 
where
\begin{equation}\label{D}
D_{\sigma\sigma'}(\hat r, {\vec p})=\xi^+_{\sigma}({\vec p})\hat r \xi_{\sigma'}({\vec p})
\end{equation}
are the usual matrix elements of the fundamental representation of the little group $SU(2)$ in the basis of polarization spinors $\{\xi\}$. The rotations $\hat r \in SU(2)$  transform this  basis and implicitly the mode spinors as
\begin{eqnarray}
	\hat r\, \xi_{\sigma}({\vec p})&=&\sum_{\sigma'}\xi_{\sigma'}({\vec p})D_{\sigma'\sigma}(\hat r, {\vec p})   \Rightarrow     	U_{{\vec p},\hat r \xi_{\sigma}}(x_c)=\sum_{\sigma'} U_{{\vec p},\xi_{\sigma'}}(x_c)D_{\sigma'\sigma}(\hat r, {\vec p})\,,\label{Uxi}\\
	\hat r\, \eta_{\sigma}({\vec p})&=&\sum_{\sigma'}\eta_{\sigma'}({\vec p})D^*_{\sigma'\sigma}(\hat r, {\vec p})  \Rightarrow 	V_{{\vec p},\hat r \eta_{\sigma}}(x_c)=\sum_{\sigma'} V_{{\vec p},\eta_{\sigma'}}(x_c)D^*_{\sigma'\sigma}(\hat r, {\vec p})\,.~~~~\label{Veta}
\end{eqnarray}
Therefore, we obtain the desired action of the operators of the spin symmetry
\begin{eqnarray}
	[{T}^s_{\hat r}\psi_{\xi}](t_c,{\vec x}_c)=\int d^3x_c' {\frak T}^s_{\hat r}( t_c,{\vec x}_c-{\vec x}_c^{\,\prime})\psi_{\xi}(t_c,{\vec x}_c^{\,\prime})	=\psi_{\hat r\xi}(t_c,{\vec x}_c)\,,
\end{eqnarray}
in accordance with the definition (\ref{Rs}). The next step is to  substitute the mode spinors (\ref{U}) and (\ref{V}) the integral (\ref{kerT}) for deriving the Fourier transforms
\begin{eqnarray}
	T^s_{\hat r}(t_c,{\vec p})={\cal U}_p(t_c)d({\vec p})\,  r\, \pi_+d({\vec p}){\cal U}^*_p(t_c) +{\cal V}_p(t_c)d(-{\vec p})\,  r\, \pi_-d(-{\vec p}){\cal V}^*_p(t_c)\,,
\end{eqnarray}
of the operators $T^s_{\hat r}$ in terms of projection operators, (\ref{pip}) and (\ref{pim}), and rotations (\ref{r0}).  

Hereby it results that the components of the spin operator (\ref{Spipi}) are Fourier operators whose Fourier transforms read
\begin{eqnarray}\label{Si}
			S_i(t_c,{\vec p})={\cal U}_p(t_c)d({\vec p})\,  s_i\, \pi_+d({\vec p}){\cal U}^*_p(t_c) +{\cal V}_p(t_c)d(-{\vec p})\,  s_i\, \pi_-d(-{\vec p}){\cal V}^*_p(t_c)\,,
\end{eqnarray}
where $s_i$ are the diagonal matrices (\ref{si}) which commutes with $\pi_{\pm}$. Furthermore, taking into account that the time modulation functions satisfy the conditions (\ref{VU}) and(\ref{uuvv}) we obtain the definitive form of these Fourier transforms as
\begin{eqnarray}
	S_i(t_c,{\vec p})&=&\left[|u^+(t_c,p)|^2-|u^-(t_c,p)|^2\right] s_i+\frac{2|u^-(t_c,p)|^2}{p^2} p^i {\vec p}\cdot{\vec s}\nonumber\\
	&+&\frac{i}{p}\left[   u^+(t_c,p)u^-(t_c,p)^*\pi_+ + u^+(t_c,p)^*u^-(t_c,p)\pi_-\right] \epsilon_{ijk}p^j\gamma^k  \,.~~~~\label{Spinf}
\end{eqnarray}
If, in addition, we set the rest frame vacuum imposing the limit (\ref{rfv}) then we obtain the correct limit in rest frames 
\begin{equation}
	\lim_{{\vec p}\to 0}	S_i(t_c,{\vec p})=s_i\,.
\end{equation}
In the particular case of the Minkowski spacetime, after substituting  the time modulation functions (\ref{upmM}) in Eq. (\ref{Spinf}), we obtain  the components of Pryce's spin operator \cite{B}, 
\begin{equation}\label{Pr}
	 S_i({\vec p})_{\rm Pryce}=\frac{m}{E(p)} s_i+\frac{p^i ({\vec s}\cdot{\vec p})}{E(p)(E(p)+m)} +\frac{i}{2E(p)}\epsilon_{ijk}p^j \gamma^k \,,
\end{equation}
which are the Fourier transforms of the components of the conserved spin operator of Dirac's theory in special relativity.

The components of the spin operator ${\vec S}$ act in CR through their Fourier transforms (\ref{Spinf}) according to the general rule (\ref{Y2}). The action of these operators on the field $\psi$ can be derived as in Eq. (\ref{Y2}) applying the operators 
(\ref{Si}) on the mode spinors (\ref{U}) and (\ref{V}) and using the obvious identities
\begin{eqnarray}
	\pi_+ d(\pm{\vec p}){\cal U}^*_p(t_c){\cal U}_p(t_c)d(\pm{\vec p})\pi_+&=&\pi_+\,,\label{idi1}\\
		\pi_- d({\pm\vec p}){\cal V}^*_p(t_c){\cal V}_p(t_c)d(\pm{\vec p})\pi_-&=&\pi_-\,,\label{idi2}
\end{eqnarray}
resulted from the condition (\ref{uuvv}). We obtain thus 
\begin{eqnarray}
	(S_i U_{{\vec p},\xi_{\sigma}})(x_c)&=&S_i(t_c,{\vec p})U_{{\vec p},\xi_{\sigma}}(x_c)=U_{{\vec p},\hat s_i\xi_{\sigma}}(x_c)\,,\\
	(S_i V_{{\vec p},\eta_{\sigma}})(x_c)&=&S_i(t_c,-{\vec p})V_{{\vec p},\eta_{\sigma}}(x_c)=V_{{\vec p},\hat s_i\eta_{\sigma}}(x_c)\,,	
\end{eqnarray}
recovering the desired action (\ref{Spipi}). Moreover, we verify that the operators $S_i$ are conserved as their Fourier transforms (\ref{Spinf}) satisfy the condition (\ref{cons}) in spite of their complicated dependence on time. In addition, these operators are translation invariant obeying  $\left[P^j,S_i\right]=0$ and have the expected algebraic properties 
\begin{eqnarray}
	\left[S_i(t_c,{\vec p})  ,	S_j(t_c,{\vec p})\right]=i\epsilon_{ijk}	 S_k(t_c,{\vec p})&~\Rightarrow ~&
	\left[ S_i  ,	 S_j\right]=i\epsilon_{ijk} S_k\,,\\
	{\vec S}^{\,2}(t_c,{\vec p})=\frac{3}{4}\cdot 1\in\rho_D &~\Rightarrow ~& \vec{S}^{\,2}=\frac{3}{4}\cdot 1\in\rho_D\,,
\end{eqnarray}
showing that they generate a fundamental representation of the $SU(2)$ group.

With the above elements  we may define the operator of fermion polarization  for any peculiar polarization given by a  pair of related spinors $\xi_{\sigma}({\vec p})$  and $\eta_{\sigma}({\vec p})$ assumed to satisfy the eigenvalues problems
\begin{equation}\label{snpp}
	\hat s_i  {n}^i({\vec p})\xi_{\sigma}({\vec p})	=\sigma\, \xi_{\sigma}({\vec p}) \Rightarrow 
	\hat s_i  {n}^i({\vec p})\eta_{\sigma}({\vec p})	=-\sigma\, \eta_{\sigma}({\vec p}),
\end{equation}
where the unit vector ${\vec n}({\vec p})$ gives the peculiar direction with respect to which the peculiar polarization is measured.  We define the polarization operator as the Fourier operator $W_s$ having the Fourier transform 
\begin{eqnarray}
	W_s(t_c,{\vec p})={\cal U}_p(t_c)d({\vec p})\, w({\vec p})\, \pi_+d({\vec p}){\cal U}^*_p(t_c) +{\cal V}_p(t_c)d(-{\vec p})\,  w(-{\vec p})\, \pi_-d(-{\vec p}){\cal V}^*_p(t_c)\,	\label{Pol}\,,
\end{eqnarray}
where $w({\vec p})=  s_i n^i({\vec p}) $.	This operator is conserved and translation invariant acting as 
\begin{eqnarray}
	(W_s U_{{\vec p},\xi_{\sigma}({\vec p})})(x)=W_s({\vec p})U_{{\vec p},\xi_{\sigma}({\vec p})}(x)&=&U_{{\vec p},w({\vec p})\xi_{\sigma}({\vec p})}(x)=\sigma U_{{\vec p},\xi_{\sigma}({\vec p})}(x)\,,\label{WU}\\
	(W_s V_{{\vec p},\eta_{\sigma}({\vec p})})(x)= W_s(-{\vec p})V_{{\vec p},\eta_{\sigma}({\vec p})}(x)&=&V_{{\vec p},w({\vec p}) \eta_{\sigma}({\vec p})}(x)= -\sigma V_{{\vec p},\eta_{\sigma}({\vec p})}(x)\,.~~~~~\label{WV}	
\end{eqnarray}
on the mode spinors depending on the polarization spinors satisfying Eqs. (\ref{snpp}). 

The above calculations in active mode are tedious and less intuitive because of the complicated Fourier transforms of the principal operators. This explain why the Pryce spin operator proposed long time ago was ignored for more than seven decades. The solution we proposed recently is to consider the passive mode in which we focus on the operators acting on the wave spinors (\ref{alpha}).

\section{Operators in passive mode}

In the passive mode we relate the operators acting on the free field (\ref{Psi}) in CR to corresponding operators acting on the wave spinors (\ref{alpha}) we call the {\em associated} operators. We associate thus  to  each operator $A:{\cal F}\to {\cal F}$  the pair of operators, $\tilde A: \tilde{\cal F}^+\to \tilde{\cal F}$ and $\tilde A^c: \tilde{\cal F}^-\to \tilde{\cal F}$, obeying
\begin{eqnarray}
	(A\psi)(x)=\int d^3p \sum_{\sigma}\left[U_{{\vec p},\sigma}(x) (\tilde A\alpha)_{\sigma}({\vec p})+V_{{\vec p},\sigma}(x) (\tilde A^c\beta)^ {*}_{ \sigma}({\vec p})\right]\,,\label{AAAc}
\end{eqnarray}
such that the brackets  of $A$ for two different fields, $\psi$ and $\psi'$,  can be calculated as
\begin{equation}\label{expA}
	\langle \psi, A\psi'\rangle_D=\langle \alpha, \tilde A \alpha'\rangle+\langle \beta, \tilde A^{c\, +} \beta' \rangle\,.
\end{equation}
Hereby we deduce that if $A=A^+$ is Hermitian with respect to the Dirac scalar product (\ref{sp}) then the associated operators are Hermitian with respect to the scalar product (\ref{spa}),  $\tilde A={\tilde A}^+$ and $\tilde A^c={\tilde A^c\,}^+$. For simplicity we denote the Hermitian conjugation of the operators acting on ${\cal F}$ and $\tilde{\cal F}$ with the same symbol but bearing in mind that the scalar products of these spaces are different. Note that,  the operators $A\in E[t_c]$ and their associated operators, $(\tilde A,\tilde A^c)$, may depend on time such that we must be careful considering the entire algebra we manipulate as frozen at a fixed time $t_c$. 

The new operators $\tilde A$ and $\tilde A^c$ are well-defined as their action can be derived by applying the inversion formulas (\ref{inv}) at any given instant $t_c$. We find thus  that $\tilde A$ and $\tilde A^c$ are integral operators that may depend on time acting as
\begin{eqnarray}
	\left.(\tilde A\alpha)_{\sigma}({\vec p})\right|_{t_c}&=&\int d^3p'\sum_{\sigma'}\left.\langle U_{{\vec p},\sigma},AU_{{\vec p}\,',\sigma'}\rangle_D\right|_{t_c} \alpha_{\sigma'}({\vec p}\,')\nonumber\\
	&+&\int d^3p'\sum_{\sigma'}\left.\langle U_{{\vec p},\sigma},AV_{{\vec p}\,',\sigma'}\rangle_D\right|_{t_c}  \beta^*_{\sigma'}({\vec p}\,')	\,,\label{Aab}\\
	\left.(\tilde A^c\beta)_{\sigma}({\vec p})\right|_t&=&\int d^3p'\sum_{\sigma'}\left.\langle U_{{\vec p}\,',\sigma'},AV_{{\vec p},\sigma}\rangle_D\right|_{t_c}  \alpha^*_{\sigma'}({\vec p}\,')\nonumber\\
	&+&\int d^3p'\sum_{\sigma'}\left.\langle V_{{\vec p}\,',\sigma'},AV_{{\vec p},\sigma}\rangle_D\right|_t  \beta_{\sigma'}({\vec p}\,')	\,,\label{Acab}
\end{eqnarray}
through kernels which are the matrix elements of the operator $A$ in the basis of mode spinors.  

In what follows we restrict ourselves to {reducible} operators which do not mix the particle and antiparticle subspaces complying with the condition (\ref{red})  at any time. In the particular case of  reducible Fourier operators, $A\in\, F[t]$, having time-dependent Fourier transforms $A(t,{\vec p})$, the non-vanishing matrix elements can be calculated easier as 
\begin{eqnarray}
&&	\left.\langle U_{{\vec p},\sigma},AU_{{\vec p}\,',\sigma'}\rangle_D\right|_{t_c} =\left.\langle U_{{\vec p},\sigma},A(t_c,{\vec p}\,')U_{{\vec p}\,',\sigma'}\rangle_D\right|_{t_c}\nonumber\\
	&&~~~~~~=\delta^3({\vec p}-{\vec p}\,')\mathring{u}_{\sigma}^+({\vec p})\pi_+d({\vec p}){\cal U}^*_p(t_c)A(t_c,{\vec p}){\cal U}_p(t_c)d({\vec p})\pi_+\,\mathring{u}_{\sigma'}({\vec p})	\,,\label{mat1}\\
		&&\left.\langle V_{{\vec p},\sigma},AV_{{\vec p}\,',\sigma'}\rangle_D\right|_{t_c} =\left.\langle V_{{\vec p},\sigma},A(t_c,-{\vec p}\,')V_{{\vec p}\,',\sigma'}\rangle_D\right|_{t_c}\nonumber\\
	&&~~~~~~=\delta^3({\vec p}-{\vec p}\,')\mathring{v}_{\sigma}^+({\vec p})\pi_-d({\vec p}){\cal V}_p^*(t_c)A(t_c,-{\vec p}){\cal V}_p(t_c)d({\vec p})\pi_-\,\mathring{v}_{\sigma'}({\vec p})	\,,\label{mat2}
\end{eqnarray}
observing that in this case the associated operators are simple $2\times 2$ matrix-operators acting separately on the spaces $\tilde{\cal F}^+$ and  $\tilde{\cal F}^-$. We observe that  when $A(t_c,{\vec p})$ commutes with ${\cal U}_p(t_c)$, ${\cal V}_p(t_c)$, $d({\vec p})$ and $\pi_{\pm}$ then we must use directly the identities (\ref{idi1}) and (\ref{idi2}). Applying first this rule to the components  of our conserved spin operator we obtain the associated operators 
\begin{eqnarray}
	S_i~~\Rightarrow~~ \tilde S_i=-\tilde S_i^c=\frac{1}{2}\Sigma_i({\vec p}) \label{tilS}	\,,
\end{eqnarray}
where the $2\times 2$ matrices $\Sigma_i({\vec p})$ have the matrix elements 
\begin{equation}\label{Dxx}
	\Sigma_{i\,\sigma\sigma'}({\vec p})=2\mathring{u}_{\sigma}^+({\vec p})s_i \mathring{u}_{\sigma'}({\vec p})=\xi^+_{\sigma}({\vec p})\sigma_i\,\xi_{\sigma'}({\vec p})\,,	
\end{equation}
depending on the polarization spinors and having the same algebraic properties as the Pauli matrices. A similar procedure gives the simple associated operators of the polarization operator (\ref{Pol}), 
\begin{equation}
	W_s~\Rightarrow~~\tilde W_s=-\tilde W_s^c=\frac{1}{2}\sigma_3\,,\label{tilW}	
\end{equation}
according to the definition (\ref{snpp}) of the polarization spinors.

It is interesting now to see how the spin operator is related to the generators of the $E(3)$  isometries of the spacetimes $M(a)$.  As the covariant representation ${T}$ defined by Eq. (\ref{TAa})  is reducible this may be associated to a  pair of  representations whose operators $\tilde { T}\in {\rm Aut}(\tilde{\cal F}^+)$ and $\tilde {T}^c\in {\rm Aut}(\tilde{\cal F}^-)$  are related to $T$ as,
\begin{eqnarray}
	&&({T}_{r,{\vec a}}\,\psi)(x_c)\nonumber\\
	&&\hspace*{12mm}=\int d^3p \sum_{\sigma}\left[U_{{\vec p},\sigma}(x_c)(\tilde { T}_{r,{\vec a}}\, \alpha)_{\sigma}({\vec p})+V_{{\vec p},\sigma}(x_c) (\tilde { T}^c_{r,{\vec a}}\,\beta)^ {*} _{ \sigma}({\vec p})\right]\,.\label{basic}
\end{eqnarray}
In other respects, by using  the identity $(R{\vec x} )\cdot {\vec p}={\vec x}\cdot (R^{-1}{\vec p})$  we  expand Eq.  (\ref{TAa}) changing the integration variable  as   
\begin{eqnarray}
&&	({T}_{r,{\vec a}}\psi  )(t_c,{\vec x}_c)
	= r \psi  \left(t_c, R(\hat r)^{-1}({\vec x}_c-{\vec a})\right)\nonumber\\
	&&\hspace*{10mm}=\int d^3p \sum_{\sigma} \left[ r U'_{{\vec p},\sigma}(x_c)\alpha_{\sigma}({\vec p}^{\,\prime})e^{i {\vec a}\cdot{\vec p}}+ r V'_{{\vec p},\sigma}(x_c)\beta^*_{\sigma}({\vec p}^{\,\prime})e^{-i {\vec a}\cdot{\vec p}}\right]\,,\label{subT}
\end{eqnarray}
where  the new mode spinors
\begin{eqnarray}
	U'_{\vec p,\sigma}({x}_c)&=&[2\pi a(t_c)]^{-\frac{3}{2}}{e^{i{\vec p}\cdot{\vec x}_c}}{\cal U}_p(t_c)d({\vec p}\,')u_{\sigma}({\vec p}\,')\,,\label{Ucuc}\\
	V'_{\vec p,\sigma}({x}_c)&=&[2\pi a(t_c)]^{-\frac{3}{2}}{e^{-i{\vec p}\cdot{\vec x}_c}}{\cal V}_p(t_c)d({\vec p}\,') v_{\sigma}({\vec p}\,')\,,\label{Vcuc}
\end{eqnarray}
depend on the transformed momentum
\begin{equation}\label{pLp}
	{{\vec p}\,}'=R(\hat r)^{-1}{\vec p}~~\Rightarrow~~ |{\vec p}\,'|=|{\vec p}|\,.	
\end{equation}
Observing then that, according to Eq. (\ref{canh}), we have $r d({\vec p}\,')= d({\vec p}) r$ we deduce that   $\tilde {T}_{r,{\vec a}} \simeq \tilde {T}^c_{r,{\vec a}}$ acting alike on the subspaces  $\tilde{\cal F}^+$ and  $\tilde{\cal F}^-$ as ,
\begin{eqnarray}
	(\tilde {T}_{r,{\vec a}}\, \alpha)_{\sigma}({\vec p})=e^{i {\vec a}\cdot{\vec p}}\sum_{\sigma'}{D}_{\sigma\sigma'}(\hat r,{\vec p}) \alpha_{\sigma'}({{\vec p}\,}') \,,
	\label{Wig}
\end{eqnarray}
and similarly for $\beta$,  where ${D}(\hat r,{\vec p})$  is just the matrix (\ref{D}).

The representation  $\tilde {T}_{r,{\vec a}}$ is unitary with respect to the scalar product (\ref{spa}), $	\langle \tilde {T}_{\lambda,a} \alpha, \tilde {T}_{\lambda,a} \alpha'\rangle =\langle  \alpha,  \alpha'\rangle$.  As the covariant representations are unitary with respect to the scalar product (\ref{spD}) which can be decomposed as in Eq. (\ref{spp}) we conclude that the expansion (\ref{Psi}) establishes the unitary equivalence, ${T}  =\tilde {T}  \oplus \tilde {T} $. Consequently, the self-adjoint generators $\tilde X\in {\rm Lie}(\tilde {T})$   defined as
\begin{eqnarray}
	\tilde	P^{i}=-\left.i\frac{\partial \tilde {T}_{1,{\vec a}}}{\partial a^{i}}\right|_{{\vec a}=0}\,, \quad 
	\tilde	J_{i}=\left.i\frac{\partial \tilde {T}_{r(\theta),0}}{\partial \theta^{i}}\right|_{\theta=0}\,,
\end{eqnarray}
are related to the corresponding ones, $X\in {\rm Lie}({T})$, such that
\begin{eqnarray}\label{basicx}
	(X\psi)(x_c)&=&\int d^3p \sum_{\sigma}\left[U_{{\vec p},\sigma}(x_c)(\tilde X\, \alpha)_{\sigma}({\vec p})- \,V_{{\vec p},\sigma}(x_c) (\tilde X\,\beta)^ {*} _{ \sigma}({\vec p})\right]\,,
\end{eqnarray}
as we deduce deriving Eq. (\ref{basic}) with respect to the corresponding group parameter $\zeta\in (\theta^i, a^i)$ in $\zeta=0$. We find thus that the isometry generators are reducible on $\tilde{\cal F}$ obeying  $\tilde X^c=-\tilde X$ as a consequence of the fact that $\tilde {T}^c\simeq\tilde {T}$. 

The associated Abelian generators are trivial being diagonal in momentum basis,
\begin{equation}\label{tilHP}
P^i ~~\Rightarrow~~ 	 \tilde P^i=-\tilde P^{c\,i}=p^i\,.
\end{equation} 
For rotations we use the Cayley-Klein parameters as in Eq. (\ref{r}) recovering the natural splitting (\ref{spli}),   
\begin{eqnarray}\label{tilJ}
	J_i=L_i+S_i~~&\Rightarrow&~~ \tilde J_i=-\tilde J^c_i=\tilde L_i+\tilde S_i\,,
\end{eqnarray}
laying out the components of the spin operator (\ref{tilS})  and intuitive components of the orbital angular momentum operator,
\begin{equation}\label{tilL}
	L_i~~\Rightarrow~~\tilde L_i =-\tilde L_i^c=-i\epsilon_{ijk}p^j\tilde \partial_k\,,
\end{equation}
where $\tilde \partial_k$ are the  {\em covariant} derivatives \cite{Cot},
\begin{equation}\label{covD}
	\tilde\partial_i=\partial_{p^i} 1_{2\times 2}+\Omega_i({\vec p})\,,
\end{equation}
 defined such that $\tilde\partial_i [\xi_{\sigma}({\vec p})\alpha_{\sigma}({\vec p})] =\xi_{\sigma}({\vec p})\tilde\partial_i \alpha_{\sigma}({\vec p}$. Therefore, the connections, 
\begin{eqnarray}
	\Omega_{i\,\sigma\sigma'}({\vec p})=\xi^+_{\sigma}({\vec p})\left[\partial_{p^i}\xi_{\sigma'}({\vec p})\right] 
	=\left\{\eta^+_{\sigma}({\vec p})\left[\partial_{p^i}\eta_{\sigma'}({\vec p})\right]\right\}^*\,,\label{Omega}
\end{eqnarray}
are anti-Hermitian,   $\Omega_{i\,\sigma\sigma'}({\vec p})=-\Omega_{i\,\sigma'\sigma}^*({\vec p})$, such that the operators $i\tilde \partial_i$ are Hermitian.  The principal property of the covariant derivatives is to commute with the spin components, $[\tilde\partial_i, \tilde S_j]=0$ thanks to  the connections $\Omega_i({\vec p})$. This becomes trivial in the case of common polarization when $\Omega_i=0$ and $\tilde S_i$ are independent on ${\vec p}$. Note that we proposed these derivatives for the first time in  Dirac's theory in Minkowski spacetime \cite{Cot}. 

The sets of conserved operators $\{\tilde L_1,\tilde L_2,\tilde L_3\}$ and $\{\tilde S_1,\tilde S_2,\tilde S_3\}$  generate the representations $\tilde T^o$ and $\tilde T^s$ of the associated factorization 
\begin{equation}\label{factor}
	{T}^r=	T^o\otimes T^s~~ \Rightarrow~~\tilde{T}^r=	\tilde T^o\otimes \tilde T^s\,,
\end{equation}	
of the $SU(2)$ restriction $\tilde{T}^r\equiv\left.\tilde{T}\right|_{SU(2)}$ of the representation $\tilde{T}$. Thus we have found the generators of the associated orbital representation studying the  isometry generators without resorting to new spectral representations. In this manner we cannot come back to the active mode in CR but we have all we need for performing the quantization. 

\section{Quantization}

 The association between the operator acting  in CR  and MR allows us to derive at any time the expectation values of  operators defined in MR according to the general rule (\ref{expA}).  We may apply thus the Bogolyubov method \cite{Bog} for quantizing the  isometry generators of the massive Dirac fermions of  arbitrary polarization.   According to this method,   we replace first  the wave spinors in MR with field operators, $(\alpha, \alpha^*)\to ({\frak a},{\frak a}^{\dag})$ and $(\beta, \beta^*)\to ({\frak b},{\frak b}^{\dag})$,  satisfying  canonical   anti-commutation relations among them  the non-vanishing ones are, 
\begin{eqnarray}\label{cac}
	\left\{{\frak a}_{\sigma}({\vec p}),{\frak a}_{\sigma'}^{\dag}({\vec p}^{\,\prime})\right\}=	\left\{{\frak b}_{\sigma}({\vec p}),{\frak b}_{\sigma'}^{\dag}({\vec p}^{\,\prime})\right\}=\delta_{\sigma\sigma'}\delta^3({\vec p}-{\vec p}^{\,\prime})\,.
\end{eqnarray}
The Dirac free field becomes thus the field operator  
\begin{eqnarray}\label{Psiq}
	\psi(x)=\int d^3p \sum_{\sigma}\left[U_{{\vec p},\sigma}(x) {\frak a}_{\sigma}({\vec p}) +V_{{\vec p},\sigma}(x) {\frak b}^ {\dag} _{ \sigma}({\vec p})\right]\,,
\end{eqnarray}  
denoted with the same symbol but acting on  the Fock state space equipped with the scalar product $\langle~~|~~\rangle$ and a normalized vacuum state $|0\rangle$ accomplishing
\begin{equation}
	{\frak a}_{\sigma}({\vec p})|0\rangle={\frak b}_{\sigma}({\vec p})|0\rangle=0\,,\quad \langle 0|{\frak a}_{\sigma}^{\dagger}({\vec
		p})=\langle 0|  {\frak b}_{\sigma}^{\dagger}({\vec p})=0\,.
\end{equation}
The sectors with different number of particles have to be constructed applying the standard method for constructing generalized momentum bases of various polarizations.

Through quantization the  expectation value of any  time-dependent operator $A(t)$  of RQM becomes  a one-particle operator 
\begin{equation}\label{qA} 
	A(t_c)~\to ~ \mathsf{A}=\left.:\langle\psi , A(t_c)\psi\rangle_D :\right| _{t_c=t_{c\,0}}\,,
\end{equation}
calculated  respecting the normal ordering of the operator products \cite{BDR,KH} at the initial time $t_{c\,0}$.  This procedure allows us to write down any one-particle operator $\mathsf{A}$ directly in terms of operators $(\tilde A,\tilde A^c)$   associated to the operator $A=A(t_c)|_{t_{c\,0}}$. We consider  here only the reducible operators  for which we  obtain the  general formula
\begin{equation}\label{Aq}
	\mathsf{A}=\int d^3{p} \left[ {\frak a}^{\dag}({\vec p})(\tilde A {\frak a})({\vec p}) -  {\frak b}^{\dag}({\vec p})(\tilde A^{c\,+} {\frak b})({\vec p})\right]\,,
\end{equation}
written with the compact notation
\begin{equation}\label{compnot}
	{\frak a}^{\dag}({\vec p})(\tilde A {\frak a})({\vec p})\equiv\sum_{\sigma} {\frak a}^{\dag}_{\sigma}({\vec p})(\tilde A {\frak a})_{\sigma}({\vec p})\,,
\end{equation}
and similarly for the second term.  We specify that the bracket (\ref{qA}) is calculated according to Eq. (\ref{expA}) in which the last term changes its sign after introducing the normal ordering of the operator products. 

Given an arbitrary operator $A\in {\rm Aut}({\cal F})$ and its Hermitian conjugated $A^+$ we define the adjoint operator of $\mathsf{A}$, 
\begin{eqnarray}
	A^+(t_c)~\to ~ \mathsf{A}^{\dagger}=\left.:\langle \psi , A(t)^+\psi\rangle_D :\right| _{t_{c\,0}}=\left.:\langle A(t) \psi , \psi\rangle_D :\right| _{t_{c\,0}}\,,
\end{eqnarray}
complying with the standard definition  $\langle \alpha |\mathsf{A}^{\dagger}\beta\rangle =\langle\mathsf{A} \alpha |\beta\rangle$ on the Fock space.  In what follows we  meet only  self-adjoint one-particle operators  as all their corresponding operators of RQM are reducible and Hermitian with respect to the scalar products of the spaces in which they act. We obtain thus an  operator algebra  formed by fields  and self-adjoint one-particle operators which have the obvious properties
\begin{eqnarray}
	\left[\mathsf{A}, \psi(x)\right]=-(A\psi)(x)\,,\qquad
	\left[\mathsf{A}, \mathsf{B}\right]=:\left<\psi, [A,B]\psi\right>_D: \,,\label{algXX1}
\end{eqnarray} 
preserving the structures of Lie algebras but without taking over other algebraic properties of their parent operators from  RQM as the product of two one-particle operators is no longer an operator of the same type. Therefore, we must restrict ourselves to the Lie algebras of symmetry generators  and unitary transformations  whose actions reduce to sums of successive commutations.  

The simplest one-particle operator is the electric charge 
\begin{equation}
	\mathsf{Q}=:\langle\psi, \psi\rangle_D:=\int d^3p\,\left[{\frak a}^{\dag}({\vec p}){\frak a}({\vec p}) -{\frak  b}^{\dag}({\vec p}){\frak b}({\vec p})\right]\,,\label{Q} 	
\end{equation}
related to the internal gauge symmetry $U(1)_{\rm em}$ \cite{BDR}. 
Other diagonal operators in momentum basis are the momentum  components  
\begin{eqnarray}
	\mathsf{P}^i&=&:\langle\psi, P^i\psi\rangle_D:=\int d^3p\,p^i\left[{\frak a}^{\dag}({\vec p}){\frak a}({\vec p}) +{\frak b}^{\dag}({\vec p}){\frak b}({\vec p})\right]\,, \label{Pom}	
\end{eqnarray}
as well as our new operator of fermion polarization ,	
\begin{eqnarray}
	\mathsf{W}_s =:\langle \psi, W_s\psi\rangle_D: =\frac{1}{2}\int d^3p\left[{\frak a}^{\dag}({\vec p}){\sigma}_3
	{\frak a}({\vec p}) +{\frak b}^{\dag}({\vec p}){\sigma}_3{\frak b}({\vec p})\right]\,, \label{Polq}
\end{eqnarray}
which enters in the incomplete set of commuting operators 
$\{\mathsf{P}^1,\mathsf{P}^2, \mathsf{P}^3,\mathsf{W}_s,\mathsf{Q}\}$	
determining the momentum bases of the Fock state space up to an integration constant that has to be fixed setting the vacuum as in the case of our rest frame vacuum defined by Eqs. (\ref{rfv}). In Minkowski spacetime we do not have this inconvenience as there exists a conserved  energy operator able to complete this set commuting with all the other operators.   In contrast, in the de Sitter expanding universe we have a conserved energy operator but this does not commute with the momentum components.  Thus the problem of setting the vacuum arises in all the FLRW spacetimes apart from the Minkowski one.   

Furthermore, applying the general rule (\ref{Aq})  to the rotation generators we find first the splitting of the total angular momentum 
\begin{equation}\label{split}
	\mathsf{J}_i=:\langle\psi, J_i\psi\rangle_D:=\mathsf{L}_i+\mathsf{S}_i\,,
\end{equation} 
where  the components of the orbital angular momentum, $\mathsf{L}_i$, and spin operator, $\mathsf{S}_i$, can be written as  
\begin{eqnarray}
	\mathsf{L}_i &=&-\frac{i}{2}\int d^3p\, \epsilon_{ijk} p^j \left[{a}^{\dag}({\vec p}){\stackrel{\leftrightarrow}{\tilde\partial_{i}}}{a}({\vec p})+{b}^{\dag}({\vec p}){\stackrel{\leftrightarrow}{\tilde\partial_{i}}}{b}({\vec p})\right]\,,\label{Lang}\\
	\mathsf{S}_i &=&\frac{1}{2}\int d^3p\left[{a}^{\dag}({\vec p})\Sigma_{i}({\vec p}){a}({\vec p})+{b}^{\dag}({\vec p})\Sigma_{i}({\vec p}){b}({\vec p})\right]\,,
	\label{Spin}
\end{eqnarray}
according to Eqs. (\ref{tilL}) and (\ref{tilS}),  using the special notation
\begin{equation}\label{stak}
	\alpha^+ \stackrel{\leftrightarrow}{\tilde\partial_{i}} \beta =\alpha^+ (\partial_{p^i}\beta)-(\partial_{p^i}\alpha^+)\beta +2\alpha^+\Omega({\vec p}) \beta\,,
\end{equation}  
inspired by Green's theorem, which points out explicitly that $\mathsf{L}_i$ are self-adjoint operators. The components $\mathsf{L}_i $ and  $\mathsf{S}_i$  form the bases of two {independent} unitary representations of the  $su(2)\sim so(3)$ algebra,  $\left[\mathsf{L}_i,\mathsf{S}_j\right]=0$, generating the orbital and respectively spin symmetries. Moreover, these operators are {conserved}  while the commutation relations 
\begin{equation}
	\left[\mathsf{L}_i,\mathsf{P}^j\right]=i\epsilon_{ijk}\mathsf{P}^k\,, \qquad 	\left[\mathsf{S}_i,\mathsf{P}^j\right]=0\,,
\end{equation}
show that only the spin operator is, in addition,   invariant under space translations.  

We obtained thus all the one-particle operators of QFT coming from the conserved operators of RQM. All these operators have similar forms in any spacetime $M(a)$, including the Minkowski one,  being independent on the scale factors. However, this universality is limited as the evolution of  the time-dependent operators is determined by the 
time modulation functions satisfying the system (\ref{sy2c}).  For example, if we apply the above procedures to the Hamiltonian (\ref{Hc}) at an arbitrary instant $t_c$ we obtain the time-dependent operator 
\begin{eqnarray}
	\mathsf{H}_c(t_c)=\left.:\langle\psi, H_c(t_c)\psi\rangle_D:\right|_{t_c}=\int d^3p\,\tilde H_c(t_c,{p})\left[{\frak a}^{\dag}({\vec p}){\frak a}({\vec p}) +{\frak b}^{\dag}({\vec p}){\frak b}({\vec p})\right]\,, 	
\end{eqnarray} 
where the quantity 
\begin{eqnarray}
H_c(t_c,{p})&=&	m\,a(t_c)\left( |u^+(t_c,p)|^2- |u^-(t_c,p)|^2 \right) \nonumber\\
&+&p\left( u^+(t_c,p)u^-(t_c,p)^*+u^+(t_c,p)^*u^-(t_c,p)\right)-\frac{3i}{2} \frac{\dot a(t_c)}{a(t_c)}\,,~~~~
\end{eqnarray}
plays the role of energy giving just the special relativistic energy $E(p)$ in the flat limit when $a\to 1$, $\dot a\to 0$ and the time modulation functions take the form (\ref{upmM}). In other respects, the form of the orbital angular momentum (\ref{Lang}) suggests us that the operator of the initial position related to the conserved spin as in Eq. (\ref{spli}) could have the same form as in the flat case, \cite{Cot,Cotnew}
\begin{equation}
	\mathsf{X}^i=\frac{i}{2}\int d^3p\left[ {\frak a}^{\dag}({\vec p})\stackrel{\leftrightarrow}{\tilde\partial_{i}} {\frak a}({\vec p}) -{\frak b}^{\dag}({\vec p})\stackrel{\leftrightarrow}{\tilde\partial_{i}} {\frak b}({\vec p}) \right]\,.
\end{equation}
For studying the time evolution of this operator we need to calculate commutators as $[ 	\mathsf{H}_c(t_c), 	\mathsf{X}^i]$ which depends on the derivatives of  time modulation functions, $\partial_p u^{\pm}(t_c,p)$. Therefore we may conclude that the time evolution of the principal observables must be studied in each particular case separately.

\appendix

\section{ $SL(2,\mathbb{C})$ transformations}

\setcounter{equation}{0} \renewcommand{\theequation}
{A.\arabic{equation}}

The Dirac field $\psi:M\to {\cal V}_D$ takes values in the space ${\cal V}_D={\cal V}_P\oplus{\cal V}_P$ of the Dirac representation  $\rho_ D=(\frac{1}{2},0)\oplus(0,\frac{1}{2})$ of the  $SL(2,\Bbb C)$ group where one defines the Dirac  $\gamma$-matrices with local indices and the Hermitian form $\overline{\psi }\psi$ with the Dirac adjoint $\overline{\psi}=\psi^+\gamma^0$ of $\psi$. These  matrices satisfy  the anticommutation rules 
\begin{equation}\label{acom}
\{\gamma^{\hat\mu},\gamma^{\hat\nu}\}=2\eta^{\hat\mu\hat\nu}\,
\end{equation}
 giving rise to the $SL(2,\mathbb{C})$ generators
\begin{equation}\label{gen}
	s^{\hat\mu\hat\nu}=\frac{i}{4}\left[\gamma^{\hat\mu},\gamma^{\hat\nu}\right]=\overline{s^{\hat\mu\hat\nu}}\in { \rho_D}[sl(2,\mathbb{C})]
\end{equation}
which are Dirac self-adjoint such that the transformations
\begin{equation}\label{tr}
	\lambda(\hat\omega)=\exp\left(-\frac{i}{2}\hat\omega^{\hat\alpha\hat\beta}s_{\hat\alpha\hat\beta}\right)\in \rho_D[SL(2,\mathbb{C})]\,, 
\end{equation}
having  real-valued parameters, $\hat\omega^{\hat\alpha \hat\beta}=-\hat\omega^{\hat\beta\hat\alpha}$,   leave the Hermitian form invariant as $\overline{\lambda(\hat\omega)}=\lambda^{-1}(\hat\omega)=\lambda(-\hat\omega)$. The corresponding  Lorentz transformations,  
$\Lambda^{\hat\mu\,\cdot}_{\cdot\,\hat\nu}(\hat\omega)\equiv	\Lambda^{\hat\mu\,\cdot}_{\cdot\,\hat\nu}[\lambda(\hat\omega)]=\delta^{\hat\mu}_{\hat\nu}
+\hat\omega^{\hat\mu\,\cdot}_{\cdot\,\hat\nu}+\frac{1}{2}\,\hat\omega^{\hat\mu\,\cdot}_{\cdot\,\hat\alpha}\hat\omega^{\hat\alpha,\cdot}_{\cdot\,\hat\nu}+\cdots$
satisfy the identities
\begin{equation}\label{canh}
	\lambda^{-1}(\hat\omega)\gamma^{\hat\alpha}\lambda(\hat\omega)	=\Lambda(\hat\omega)^{\hat\alpha\,\cdot}_{\cdot \,\hat\beta}\gamma^{\hat\beta}\,,
\end{equation} 
which encapsulate the canonical homomorphism \cite{WKT}.

In the chiral representation of the Dirac matrices (with diagonal $\gamma^5$)  the transformations  $\lambda(\hat\omega)$  generated by the matrices $s^{\mu\nu}$ are reducible to the  subspaces of Pauli spinors ${\cal V}_P$ carrying the irreducible representations $(\frac{1}{2},0)$ and $(0,\frac{1}{2})$ of ${\rho_D}$ \cite{WKT,Th}. 
We denote  by 
\begin{equation}\label{r0}
	r={\rm diag}(\hat r,\hat r)\in {\rho_D}\left[SU(2)\right]	
\end{equation}
the transformations we call here simply rotations, and by 
\begin{equation}\label{l0}
	l={\rm diag}(\hat l,\hat l^{-1})\in {\rho_D}\left[ SL(2,\mathbb{C})/SU(2)\right]	
\end{equation}
the Lorentz boosts. For rotations we use the  generators  
\begin{equation}\label{si}
	s_i= \frac{1}{2}\epsilon_{ijk}s^{jk}  ={\rm diag}(\hat s_i,\hat s_i)\,, \quad \hat s_i=\frac{1}{2}\sigma_i\,,
\end{equation}
and Cayley-Klein parameters  $\theta^i=\frac{1}{2}\epsilon_{ijk}\hat\omega^{jk}$ such that
\begin{eqnarray}
	r(\theta)={\rm diag}(\hat r(\theta),\hat r(\theta))\,,\quad~~ \hat r(\theta)=e^{-i \theta^i \hat s_i}=e^{-\frac{i}{2} \theta^i \sigma_i}\,, \label{r}
\end{eqnarray}
where  $\sigma _i$ are the Pauli matrices.  Similarly, we chose the parameters  $\tau^i=\hat\omega^{0i}$ and the generators $s_{0i}=-s^{0i}={\rm diag}(i\hat s_i, -i\hat s_i)$ for the Lorentz boosts,  
$l(\tau)={\rm diag}(\hat l(\tau),\hat l^{-1}(\tau))$ where $ \hat l(\tau)=e^{ \tau^i \hat s_i}=e^{\frac{1}{2} \tau^i \sigma_i}$. 
The corresponding transformations of the group $L_+^{\uparrow}$ will be denoted as $R(\hat r)=\Lambda (r)$ and $L(\hat l)=\Lambda(l)$.

Particularly, the boosts (\ref{l0}) of Wigner's method of constructing the covariant Dirac field in Minkowski spacetime have the parameters $\tau^i=-\frac{p^i}{p}{\rm tanh}^{-1} \frac{p}{E(p)}$ taking the form   \cite{Th}
\begin{eqnarray}\label{Ap}
	l_{{\vec p}}=\frac{E(p)+m+\gamma^0{\vec\gamma}\cdot {\vec p}}{\sqrt{2m(E(p)+m)}}=	l_{{\vec p}}^+\,,\quad l^{-1}_{\vec p}=l_{-\vec p}=\bar{l}_{\vec p}\,,
\end{eqnarray}
where $E(p)=\sqrt{m^2+{\vec p}^2}$ is the energy in special relativity. These boosts which determine the form of the mode spinors in Minkowski spacetime can be related as 
\begin{equation}\label{id1}
	\sqrt{\frac{m}{E(p)}} l_{\vec p}\,e^{-iE(p)t}=
	 u^{+}_M(t,p)+\frac{\gamma^0{\vec\gamma}\cdot {\vec p}}{p} u_M^-(t,p)
\end{equation}
to the time modulation functions in this manifold, 
\begin{equation}\label{upmM}
u^{\pm}_M(t,p)=\sqrt{\frac{E(p)\pm m}{2E(p)}}e^{-iE(p)t}	\,.
\end{equation}
Hereby we deduce the identity 
\begin{equation}\label{id2}
	\left( u^{+}_M(t,p)+\frac{\gamma^0{\vec\gamma}\cdot {\vec p}}{p} u_M^-(t,p)  \right)\pi_+=\left(u_M^+(t,p)\pi_++u_M^-(t,p)\pi_-\right)\left( 1+ \frac{\gamma^0{\vec\gamma}\cdot {\vec p}}{p}\right)\pi_+
\end{equation}
that inspires our method of constructing Dirac mode spinors in any FLRW spacetime with the help of the matrix (\ref{gamp}) revealed here.

\end{document}